# ALGEBRAIC REPRESENTATION OF MANY-PARTICLE COULOMB GREEN FUNCTION AND APPLICATION IN ATOMIC CALCULATIONS

NGUYEN HOANG QUOC,
NGUYEN NGOC TY, LE VAN HOANG, NGUYEN KHAC NHAP

Department of Physics, HCMC University of Pedagogy
280 An Duong Vuong, 5 Dist., HCM city
Email: hoanglv@hcm.fpt.vn

**Abstract.** Basing on the relation between the Coulomb Green function and the Green function of harmonic oscillator, the algebraic representation of the many-particle Coulomb Green function in the form of annihilation and creation operators is established. These results allow us to construct purely algebraic method for atomic calculations and thus to reduce rather complicated calculations of matrix elements to algebraic procedures of transforming the products of annihilation and creation operators to a normal form.  Effectiveness of the constructed method is demonstrated through the example problem: the ground state of hydrogenic molecule. The calculation algorithm of this algebraic approach is simple and suitable for symbolic calculation programs, such as Mathematica, that widely enlarge the application field of the Coulomb Green function.

## I.  INTRODUCTION

Perturbation theory method using the Coulomb Green function is one of the most effective analytical methods for solving Schrödinger equation of atomic systems [1]. The success in constructing the algebraic representation of the Coulomb Green function [2] allows one to use purely algebraic method in calculating matrix elements of the Coulomb Green function (CGF). This algebraic approach reduces rather complicated calculations of matrix elements into the algebraic procedure of transforming products of creation and annihilation operators to a normal form (see, [2-4]) and thus considerably enlarges its application area [5-7]. One of the advantages of the algebraic method is the capability of using symbolic calculation programs, such as Mathematica for automation of calculation process. Because atomic calculations frequently lead to the enormous amount of complicated integrals so the automation has an extremely important practicality. The method described in [2], which is called the algebraic method of using Coulomb Green function, is applied fruitfully for solving the problems of a hydrogen-like atom in an electromagnetic field (see [5-6] and the topic review [7]).

In present work, we will develop the above-mentioned algebraic method for the problems of a many-electron atom or molecule in an external electromagnetic field. For this purpose, the Coulomb Green function for many Coulomb centers, called the many-particle Coulomb Green (MCGF), is successfully established in the form of multiplying CGF for each Coulomb center. For illustration of the algebraic calculation, we consider the problem of the ground state energy of a hydrogenic molecule as an example of application. The other actual problems [8-10] will be considered in our next works (see the report [11]).

## II.  ALGEBRAIC METHOD

For use in the next section of the work, we pay attention to some main points of the algebraic method given in [2-3].

One of the main results in [3] (see also [12]) is the construction of the hydrogen-like atom wavefunctions in the form of creation operators acting in the vacuum-state as in the second-quantization representation. For example, the $s$-state vectors have the form:

$$|ns\rangle = \frac{1}{\sqrt{n!(n+1)!}} (\hat{a}_t^+ \hat{b}_t^+)^n |0(\omega)\rangle, \tag{1}$$

and the p-state vector ($m=1$) is

$$|np\rangle = \frac{1}{\sqrt{n!(n+1)!}} (\hat{a}_t^+ \hat{b}_t^+)^{n-1} \hat{a}_1^+ \hat{b}_2^+ |0(\omega)\rangle. \tag{2}$$

The other state vectors can be constructed in the familiar form (see [3, 4 and 12]). Here, $n+1$ is the principal quantum number; $m$ is the azimuthal quantum number; $\omega$ is a positive real number defined later on; repeating indices summation over them ($t=1, 2$); the annihilation and creation operators used in (1) (2) have the typical commutative correlations:

$$[\hat{a}_s(\omega), \hat{a}_t^+(\omega)] = \delta_{st}, \quad [\hat{b}_s(\omega), \hat{b}_t^+(\omega)] = \delta_{st} \tag{3}$$

(the other commutations are equal to zero); the vacuum-state is defined by equations:

$$\hat{a}_s |0(\omega)\rangle = 0, \quad \hat{b}_s |0(\omega)\rangle = 0 \quad (s=1,2), \quad \langle 0(\omega)|0(\omega)\rangle = 1. \tag{4}$$

Henceforth, we use the following notations:

$$\begin{aligned} N+2 &= \hat{a}_s^+ \hat{a}_s + \hat{b}_s^+ \hat{b}_s + 2, \quad n_\lambda^a = (\sigma_\lambda)_{st} \hat{a}_s^+ \hat{a}_t, \quad n_\lambda^b = (\sigma_\lambda)_{st} \hat{b}_t^+ \hat{b}_s \\ M &= \hat{a}_s \hat{b}_s, \quad M^+ = \hat{a}_s^+ \hat{b}_s^+, \quad m_\lambda = (\sigma_\lambda)_{st} \hat{a}_t \hat{b}_s, \quad m_\lambda^+ = (\sigma_\lambda)_{st} \hat{a}_s^+ \hat{b}_t^+, \end{aligned} \tag{5}$$

with $\sigma_\lambda$ ($\lambda = 1,2,3$) are the Pauli matrices. All operators corresponding to physical quantities can be expressed via the operators in (5), for example:

$$r = \frac{1}{2\omega}(M + M^+ + N + 2), \quad x_\lambda = \frac{1}{2\omega}(m_\lambda + m_\lambda^+ + n_\lambda^a + n_\lambda^b), \quad rp_\lambda \Rightarrow -\frac{i}{2}(m_\lambda - m_\lambda^+),$$

$$r\mathbf{p}^2 = \frac{\omega}{2}(N + 2 - M - M^+), \quad \hat{L}^2 = \frac{1}{4} N(N+2) - M^+ M, \quad \hat{l}_\lambda = n_\lambda^a - n_\lambda^b. \tag{6}$$

Formulae (1)-(6) describe an extremely important result in practice because they lead all atomic calculations to purely algebraic operations without using the evident form of hydrogen-like atom wavefunctions. Meanwhile, when necessary, we can easy receive the latter from its algebraic representation (1)-(2) with the use of the following definition:

$$\begin{aligned} \hat{a}_s(\omega) &= \sqrt{\frac{\omega}{2}} \left( \xi_s + \frac{1}{\omega} \frac{\partial}{\partial \xi_s^*} \right), \quad \hat{a}_s^+(\omega) = \sqrt{\frac{\omega}{2}} \left( \xi_s^* - \frac{1}{\omega} \frac{\partial}{\partial \xi_s} \right), \\ \hat{b}_s(\omega) &= \sqrt{\frac{\omega}{2}} \left( \xi_s^* + \frac{1}{\omega} \frac{\partial}{\partial \xi_s} \right), \quad \hat{b}_s^+(\omega) = \sqrt{\frac{\omega}{2}} \left( \xi_s - \frac{1}{\omega} \frac{\partial}{\partial \xi_s^*} \right). \end{aligned} \tag{7}$$

Here $\xi_s (s=1,2)$ are coordinates in two-dimensional complex space, related to our real physical three-dimensional space via the Kustaanheimo-Stiefel transformation [3]:

$$\begin{cases} x_\lambda = \xi_s^* (\sigma_\lambda)_{st} \xi_t \\ \phi = \arg(\xi_1), \ 0 \leq \phi \leq 2\pi. \end{cases} \quad (8)$$

It has been proved that by using the transformation (8), the Schrödinger equation of hydrogen-like atom leads to the one of a harmonic oscillator [3]. Moreover, the relation between CGF and the Green function of harmonic oscillator [2] (see also [13, 14]) has been established correspondently:

$$K(\mathbf{r},\mathbf{r}';E) = \frac{1}{8} \int_0^{2\pi} d\phi' \, U(\mathbf{r},\mathbf{r}',\phi - \phi'; Ze^2) \quad (9)$$

with $K(\mathbf{r},\mathbf{r}';E)$ is CGF and $U(\mathbf{r},\mathbf{r}',\phi - \phi'; Ze^2)$ is the Green function for two-dimensional complex harmonic oscillator. By using the relation (9), the Green operators corresponding to CGF (Coulomb Green operator) can be expressed in the formula:

$$\hat{K}(\mathbf{r};E) = \hat{U}(\xi, Ze^2) \xi_s \xi_s^*. \quad (10)$$

Here $\hat{U}(\xi, Ze^2)$ is the Green operator of harmonic oscillator, which will be used in concrete calculations in the algebraic form:

$$\hat{U}(Ze^2) = -i \lim_{\varepsilon \to +0} \int_0^{+\infty} dt \, e^{-\varepsilon t + iZe^2 t} \exp\left\{ \frac{i\omega t}{2}(N+2) \right\} \quad (11)$$

if $E = -\omega^2/2$. In case of $E = \pm \nu^2/2, \nu \neq \omega \ (\nu \geq 0)$, which is often met in atomic calculations, the exponential factor in (11) must be changed to the form:

$$\exp\left\{ -it \left( \frac{-\omega^2 \pm \nu^2}{4\omega}(M+M^+) + \frac{\omega^2 \pm \nu^2}{4\omega}(N+2) \right) \right\}. \quad (12)$$

It is easy to verify that 15 operators (5) generate a closed algebra. Thus, the factor (12) can be reduced into a normal form. It means the annihilation operators locate in the right-hand side and all the creation ones locate in the left-hand side. Therefore, it is suitable for acting in the state-vector such as (1)-(2) and can be applied in concrete calculations using algebraic approach. We called the form (10) with (11) and (12) algebraic representation of CGF. For more details, one can see [2], [11].

### III. MANY-PARTICLE COULOMB GREEN FUNCTION

When considering the many-electron atom problems as well as the problem of atom or molecule interaction, we can construct Hamiltonian in two parts. The first of them is the following:

$$\hat{H} = \sum_k \hat{H}_k, \qquad \hat{H}_k = -\frac{1}{2}\Delta_k - \frac{Z_k e^2}{r_k}, \quad (k=1,2,...,N) \quad (13)$$

where $\Delta_k$ is Laplace operator corresponding to $\mathbf{r}_k$- coordinate; $Z_k$ $(k=1,2,...,N)$ are different positive numbers connected with effective charge of electrons in the independent particle model with the hydrogenic wave-functions [15], but in the conventional description of atoms they are the same and equal to nuclear charge. The remaining part of Hamiltonian $\hat{V}(\mathbf{r}_1,\mathbf{r}_2,...,\mathbf{r}_N)$ is considered as a perturbation term. The Schrödinger equation can be reduced to the integral equation:

$$\psi(\mathbf{r}_1,\mathbf{r}_2,...,\mathbf{r}_N) = -\int K(\mathbf{r}_1,\mathbf{r}_2,...,\mathbf{r}_N,\mathbf{r}'_1,\mathbf{r}'_2,...,\mathbf{r}'_N;E)\hat{V}\psi(\mathbf{r}'_1,\mathbf{r}'_2,...,\mathbf{r}'_N)d\mathbf{r}'_1 d\mathbf{r}'_2...d\mathbf{r}'_N \quad (14)$$

with using the many-particle Coulomb Green function, defined by the equation:

$$(\hat{H}-E)K(\mathbf{r}_1,\mathbf{r}_2,...,\mathbf{r}_N,\mathbf{r}'_1,\mathbf{r}'_2,...,\mathbf{r}'_N;E) = -\delta(\mathbf{r}_1-\mathbf{r}'_1)\delta(\mathbf{r}_2-\mathbf{r}'_2)\cdots\delta(\mathbf{r}_N-\mathbf{r}'_N). \quad (15)$$

Sometimes in concrete calculations, it is more convenient if we use a Green operator defined from the Green function via the following formulation:

$$\hat{K}(\mathbf{r}_1,\mathbf{r}_2,...,\mathbf{r}_N;E)\varphi(\mathbf{r}_1,\mathbf{r}_2,...,\mathbf{r}_N) =$$

$$\int K(\mathbf{r}_1,\mathbf{r}_2,...,\mathbf{r}_N,\mathbf{r}'_1,\mathbf{r}'_2,...,\mathbf{r}'_N;E)\varphi(\mathbf{r}'_1,\mathbf{r}'_2,...,\mathbf{r}'_N)d\mathbf{r}'_1 d\mathbf{r}'_2...d\mathbf{r}'_N.$$

Thus, the equation (14) becomes:

$$\psi(\mathbf{r}_1,\mathbf{r}_2,...,\mathbf{r}_N) = -\hat{K}(\mathbf{r}_1,\mathbf{r}_2,...,\mathbf{r}_N;E)\hat{V}(\mathbf{r}_1,\mathbf{r}_2,...,\mathbf{r}_N)\psi(\mathbf{r}_1,\mathbf{r}_2,...,\mathbf{r}_N).$$

For constructing the Green operator, we use $\hat{K}(\mathbf{r}_1,\mathbf{r}_2,...,\mathbf{r}_N;E) = (\hat{H}-E)^{-1}$ as a formalistic notation with the meaning that the Green operator is inverse operator of the left part of the equation (15). Using an integral representation of an inverse operator:

$$\hat{A}^{-1} = -i\lim_{\varepsilon\to+0}\int_0^{+\infty}dt\, e^{-\varepsilon t+i\hat{A}t},$$

we obtain $\quad \hat{K}(\mathbf{r}_1,\mathbf{r}_2,...,\mathbf{r}_N;E) = \dfrac{1}{i}\lim_{\varepsilon\to+0}\int_0^{+\infty}dt\, e^{-\varepsilon t+i(\hat{H}-E)t}. \quad (16)$

Now we try to express (16) in the form that is suitable to algebraic calculations via the operator of CGF of each coordinate $\mathbf{r}_k$. Accordingly, (16) can be rewritten to

$$\hat{K}(\mathbf{r}_1,\mathbf{r}_2,...,\mathbf{r}_N;E) = \frac{1}{i}\lim_{\varepsilon\to 0}\int_0^{+\infty}dt_1\int_0^{+\infty}dt_2...\int_0^{+\infty}dt_N$$

$$\times\delta(t_2-t_1)\delta(t_3-t_1)...\delta(t_N-t_1)\prod_{k=1}^{N}e^{-\varepsilon t_k+i(\hat{H}_k-E_k)t_k} \quad (17)$$

with $E = \sum_{k=1}^{N}E_k$. By substituting the Furrier integral form of delta-function into (17), we obtain:

$$\hat{K}(\mathbf{r}_1, \mathbf{r}_2, \ldots, \mathbf{r}_N; E) = \frac{1}{i(2\pi)^{N-1}} \int_{-\infty}^{+\infty} d\alpha_1 \int_{-\infty}^{+\infty} d\alpha_2 \cdots \int_{-\infty}^{+\infty} d\alpha_N$$

$$\times \delta(\alpha_1 + \alpha_2 + \ldots + \alpha_N) \lim_{\varepsilon \to 0} \int_0^{+\infty} dt_1 \int_0^{+\infty} dt_2 \cdots \int_0^{+\infty} dt_N \prod_k \exp\left\{-\varepsilon t_k + i(\hat{H}_k + \alpha_k - E_k)t_k\right\}$$

$$= \frac{1}{i(2\pi)^{N-1}} \int_{-\infty}^{+\infty} d\alpha_1 \int_{-\infty}^{+\infty} d\alpha_2 \cdots \int_{-\infty}^{+\infty} d\alpha_N \, \delta(\alpha_1 + \alpha_2 + \ldots + \alpha_N) \prod_{k=1}^{N} \hat{K}(\mathbf{r}_k; E_k - \alpha_k). \tag{18}$$

In (18), $\hat{K}(\mathbf{r}_k; E_k - \alpha_k) = \lim_{\varepsilon \to +0} (\hat{H}_k + \alpha_k - E_k + i\varepsilon)^{-1}$ is the Green operator of CGF corresponding to $\mathbf{r}_k$-coordinate and the energy $E_k - \alpha_k$. Thus, we have reduced the operator of MCGF to the separated operators of CGF. From (18), it is easy to see that because each $\hat{K}(\mathbf{r}_k; E_k - \alpha_k)$ has a normal form then $\hat{K}(\mathbf{r}_1, \mathbf{r}_2, \ldots, \mathbf{r}_N; E)$ has one either.

**Application example: Ground state of hydrogenic molecule**

The Schrödinger equation for a hydrogenic molecule has the form:

$$\left\{\frac{1}{2}\mathbf{p}_1^2 - \frac{e^2}{r_1} + \frac{1}{2}\mathbf{p}_2^2 - \frac{e^2}{r_2} + V(R, \mathbf{r}_1, \mathbf{r}_2)\right\} \Psi(\mathbf{r}_1, \mathbf{r}_2) = E(R) \Psi(\mathbf{r}_1, \mathbf{r}_2) \tag{19}$$

($\hbar = m = c = 1$) with R is the distance between two molecule nucleus; $\mathbf{r}_1, \mathbf{r}_2$ are coordinates of each electron respectively. Assuming that $R \gg r_1$, $R \gg r_2$, the potential $V(R, \mathbf{r}_1, \mathbf{r}_2)$ is expanded in the Taylor power series as follows:

$$V(\mathbf{R}, \mathbf{r}_1, \mathbf{r}_2) = (2z_1 z_2 - x_1 x_2 - y_1 y_2)\frac{e^2}{R^3}$$

$$+ \frac{3}{2}[2(z_1 z_2 + x_1 x_2 + y_1 y_2)(z_1 - z_2) + z_1(x_2^2 + y_2^2) + z_2(x_1^2 + y_1^2)]\frac{e^2}{R^4} + \cdots \tag{20}$$

For demonstration of algebraic calculation, we only pay attention to the first term in expansion (20). The other terms can be considered absolutely by the same way. The equation (19) now becomes:

$$\left\{\left(\frac{1}{2}r_1\mathbf{p}_1^2 - r_1 E_1(R) - e^2\right)r_2 + \left(\frac{1}{2}r_2\mathbf{p}_2^2 - r_2 E_2(R) - e^2\right)r_1\right.$$

$$\left. + r_1 r_2 (2z_1 z_2 - x_1 x_2 - y_1 y_2)\frac{e^2}{R^3}\right\} \Psi(\mathbf{r}_1, \mathbf{r}_2) = 0 \tag{21}$$

with $E(R) = E_1(R) + E_2(R)$. For the ground state, $E_1(R) = E_2(R) = E(R)/2$. By doing the formal change (6) for each of the coordinates, we transform (21) into an equation describing the motion of two "particles" in the space of complex coordinates ξ. Henceforth, all operators in equation (21) should be understood via the formal changes (6). Assuming the parameter $\beta = e^2/R^3$ small, we now solve equation (21) by the perturbation method. The wavefunction in zero order approximation can be acquired in the form:

$$\left|\Psi^{(0)}(\mathbf{r}_1,\mathbf{r}_2)\right\rangle = \left|\Psi(\mathbf{r}_1)\right\rangle\left|\Psi(\mathbf{r}_2)\right\rangle,$$

where $\left|\Psi(\mathbf{r})\right\rangle$ is the ground-state wave-vector of the hydrogen-like atom with energy $E^{(0)} = -\omega^2$. For the ground state, from (1) we have $\left|\Psi\right\rangle = \left|0(\omega)\right\rangle$, $\omega = e^2$.

Using the operator of two-particle CGF of the form (18), from (21), we find the first order correction to the wavefunction in ground state as follows:

$$\left|\Psi^{(1)}\right\rangle = \frac{i\beta}{2\pi} \int_{-\infty}^{+\infty} d\alpha \lim_{\varepsilon \to 0} \int_0^{+\infty} dt \exp\left\{-it\left(\frac{1}{2}r_1\mathbf{p}_1^2 + \frac{1}{2}\omega^2 r_1 - \alpha\ r_1 - e^2\right) + \varepsilon t\right\}$$

$$\times \int_0^{+\infty} ds\, \exp\left\{-is\left(\frac{1}{2}r_2\mathbf{p}_2^2 + (\frac{1}{2}\omega^2 - \alpha)r_2 - e^2\right) + \varepsilon s\right\} r_1 r_2 (2z_1 z_2 - x_1 x_2 - y_1 y_2)\left|\Psi^{(0)}\right\rangle. \quad (22)$$

Taking into account that the wavefunction in ground state depends only on the distances $r_1, r_2$ (not on the vectors $\mathbf{r}_1, \mathbf{r}_2$), we find the first order energy correction equal to zero. The second-order correction to energy can be obtained in the form:

$$E^{(2)} = \frac{3i}{\pi}\beta^2 \left\langle\Psi(\mathbf{r})|r|\Psi(\mathbf{r})\right\rangle^{-2} \int_{-\infty}^{+\infty} d\alpha \left\{\lim_{\varepsilon \to 0}\int_0^{+\infty} dt\ \exp\{ie^2 t - \varepsilon t\}\right.$$

$$\left.\times\left\langle\Psi(\mathbf{r})\right| rx \exp\left\{-it\left(\frac{1}{2}r\mathbf{p}^2 + \frac{1}{2}\omega^2 r - \alpha\ r\right)\right\} rx\left|\Psi(\mathbf{r})\right\rangle\right\}^2. \quad (23)$$

Here in (23), we eliminate the electron indices because they have no effect on the calculation of matrix elements. Putting:

$$v^2 = \frac{1}{2}\omega^2 - \alpha \quad \text{if} \quad \omega^2 \geq 2\alpha, \qquad v^2 = \alpha - \frac{1}{2}\omega^2 \quad \text{if} \quad \omega^2 < 2\alpha,$$

in (23), the exponential factor has the form like (12) which permits using the algebraic calculation. As a result, we have:

$$\left\langle 0(\omega)|r|0(\omega)\right\rangle = \frac{1}{\omega},$$

$$\left\langle 0(\omega)\right| rx \exp\left\{-it\left(\frac{1}{2}r\mathbf{p}^2 + \frac{1}{2}v^2 r\right)\right\} rx\left|0(\omega)\right\rangle =$$

$$2^{11} v^4 \omega^2 \frac{(\omega+v)^2 e^{-2ivt} + (3v^2 - 2\omega^2)e^{-3ivt} + (\omega-v)^2 e^{-4ivt}}{\left((\omega+v)^2 - (\omega-v)^2 e^{-ivt}\right)^6}. \quad (24)$$

Substituting (24) into (23) and taking integration over t and then over α, we obtain, for instance, the numerical result: $E^{(2)} = -6.499026\, e^2/R^6$ which coincides with the well-known data [16].

## IV. CONCLUSION

Within the concern of this paper, we have demonstrated the way of using operator representation of the many-particle Coulomb Green function for algebraic atomic calculations. The calculation method is simple, straightforward and thus reduces rather complicated integrations of special functions to simple, pure algebraic procedures of transforming the products of the annihilation and creation operators to a normal form.

The method can be recommended for the majority problems in atomic physics. This relates to an aspect of our further investigations.